\documentclass[useAMS,usenatbib,usegraphicx]{mn2e}

\usepackage{booktabs}
\usepackage{times}

\title[Dust growth in high-redshift quasars]
{Impact of grain size distributions on the dust enrichment
in high-redshift quasars}
\author[Tzu-Ming Kuo and Hiroyuki Hirashita]{Tzu-Ming Kuo$^{1,2}$
and Hiroyuki Hirashita$^{1}$\\
$^{1}$Institute of Astronomy and Astrophysics, Academia Sinica, PO Box 23-141, Taipei 10617, Taiwan\\
$^{2}$Department of Physics, National Taiwan University, Taipei 10617, Taiwan}
\date{2012 April 18}
\pagerange{\pageref{firstpage}--\pageref{lastpage}} \pubyear{2011}

\begin{document}
\label{firstpage}
\maketitle

\begin{abstract}
In high-redshift ($z>5$) quasars, a large amount of dust ($\textstyle\sim 10^{8}~\mathrm{M}_{\sun}$) has been observed. In order to explain the large dust content, we focus on a possibility that grain growth by the accretion of heavy elements is the dominant dust source. We adopt a chemical evolution model applicable to nearby galaxies but utilize parameters adequate to high-$z$ quasars. It is assumed that metals and dust are predominantly ejected by Type II supernovae (SNe). We have found that grain growth strongly depends on the grain size distribution. If we simply use the size distribution of grains ejected from SNe, grain growth is inefficient because of the lack of small grains (i.e.\ small surface-to-volume ratio of the dust grains). However, if we take small grain production by interstellar shattering into consideration, grain growth is efficient enough to account for the rich dust abundance in high-$z$ quasars. Our results not only confirm that grain growth is necessary to explain the large amount of dust in high-$z$ quasars, but also demonstrate that grain size distributions have a critical impact on grain growth.
\end{abstract}

\begin{keywords}
dust, extinction -- galaxies: evolution --
galaxies: high-redshift -- galaxies: ISM -- ISM: clouds --
quasars: general
\end{keywords}

\section{Introduction}

The origin of dust grains in the Universe is one of the fundamental problems in astrophysics. It is widely believed that the evolution of dust content in galaxies is influenced by dust formation in stellar ejecta, dust destruction in supernova (SN) remnants, and grain growth by the accretion of metals onto the preexisting dust grains in molecular clouds \citep[e.g.][]{dwek98}. Various authors have shown that grain growth in molecular clouds dominates the increase of dust mass in nearby galaxies {\citep{hirashita99,inoue03,zhukovska08,draine09,inoue11}}. \citet{zhukovska08} also mention that grain growth depends on the mean surface-to-volume ratio of the dust grains. In our previous work \citep[][hereafter HK11]{hirashita11}, we have investigated the effects of dust grain size distribution on grain growth, and found that the grain size distribution has a large influence on the efficiency of grain growth because the surface-to-volume ratio is governed by the grain size distribution.

The extinction curves in the Galaxy and the Magellanic Clouds
can be explained with dust grain size distributions
$n(a)\propto a^{-3.5}$, where $a$ is the grain radius and
$n(a)\,\mathrm{d}a$ is the number density of grains whose
radius is between $a$ and $a+\mathrm{d}a$
\citep{mathis77,pei92}. However, high-redshift quasars exhibit
different extinction curves from
the ones in local galaxies \citep[e.g.][]{gallerani10}, which may
indicate different grain size distributions.
{Different dust sources and/or different grain processing in
the early Universe may be the reason for the different
extinction curves in high-$z$ quasars.}
{In the early Universe, Type II supernovae (simply denoted
as SNe in this paper) should be the foremost source of dust formation because of the short lifetimes of the progenitors \citep{gall11}, albeit asymptotic giant branch (AGB) stars may have contribution even at $z>5$ for some stellar initial mass function \citep{valiante09}. Star-forming regions
may also host significant grain processing such
as shattering \citep{hirashita10}.}

Grain growth in molecular clouds may dominate the grain mass increase not only in the local Universe but also at high $z$ \citep{mattsson11,pipino11,valiante11,asano12}.
However, the potential importance of dust grain size distributions in grain growth has not been focused on. As emphasized above, the grain size distributions in high-$z$ quasars should be different from those in local galaxies. Therefore, we should reconsider grain growth by using grain size distributions applicable to high-$z$ systems.

Size distributions of grains produced and ejected from SNe (called SN-dust) are calculated by \citet{nozawa07}, who take dust condensation and destruction in SNe into account. {They show that the average grain size of SN-dust is biased to large sizes because small grains are efficiently trapped in the shocked region and considerably destroyed by thermal sputtering.} This means that the grain size distribution is completely different from the local $a^{-3.5}$-law and that the surface-to-volume ratio is smaller than the local one. As a result, we have to reexamine whether grain growth is efficient enough to explain the large dust abundance in high-$z$ quasars.

After SN-dust is injected into the interstellar medium (ISM), the grain size distribution can be modified. Dust grains acquire large velocity dispersions in a warm ionized medium through dynamical coupling with turbulent motion \citep*{yan04}. Grain--grain collisions at velocities driven by turbulence lead to efficient shattering, which produces a large number of small grains \citep{hirashita10}. The extinction curve observed in a high-$z$ quasar by \citet{maiolino04} is also reproduced by a grain size distribution after shattering \citep{hirashita10}. Shattering raises the surface-to-volume ratio, leading to a higher grain growth rate.

In this Letter, we examine the significance of grain size distribution
on grain growth in high-$z$ quasars by focusing on two different
grain size distributions as representative ones expected theoretically
in high-$z$ quasars: one is the grain size distribution of SN-dust
\citep{nozawa07}, and the other is the size distribution of grains processed by
interstellar shattering \citep{hirashita10}. This Letter is organized as follows.
In Section 2, we explain theoretical models of dust enrichment, focusing on
grain growth by accretion. In Section 3,  we show the results, which we
discuss in more general contexts in Section 4. Finally, Section 5 gives the conclusion.




\section{Model}

\subsection{Dust Enrichment}

We adopt a chemical evolution model that describes the time evolution of gas, metals, and dust in a galaxy. In order to focus on the effect of grain size distribution on grain growth, we adopt a simple model, which has already been shown to be applicable to nearby galaxies (\citealt{hirashita99}; HK11), but use parameter values suitable for high-$z$ ($z>5$) quasars. {We neglect inflow and outflow, and treat the system as a closed box.}\footnote{{As long as the inflow has much smaller abundance of metals and dust than the system, both the metallicity and the dust-to-gas ratio decrease with the same fraction. Moreover, as long as the mass loss, for instance, by mass accretion onto the central black hole (BH) or  outflow of gas heated by the feedback from BH or stars, has the same metallicity and dust-to-gas ratio as the host galaxy, it does not affect the relation between metallicity and dust-to-gas ratio.}} The abundances of dust and metals are tightly related, so the model equations are finally reduced to the relation between dust-to-gas ratio and metallicity by adopting the instantaneous recycling approximation (HK11), {which can be applied as long as we consider time-scales longer than the lifetime of massive stars}:
	\begin{eqnarray}
		\mathcal{Y}_\mathrm{X}\frac{\mathrm{d}\mathcal{D}_\mathrm{X}}{\mathrm{d}Z_\mathrm{X}} & = &
		f_\mathrm{in, X}(\mathcal{R}Z_\mathrm{X}+\mathcal{Y}_\mathrm{X})
		-(\beta_\mathrm{SN}+\mathcal{R})\mathcal{D}_\mathrm{X}\nonumber\\
		& & +\frac{1}{\psi}\left[\frac{\mathrm{d}M_\mathrm{dust, X}}{\mathrm{d}t}\right]_{\rm{acc}},\label{eq:dDdX}
	\end{eqnarray}
	where
		$Z_\mathrm{X}$ is the mass abundance of element X (X is a key element
		of a dust species; X = Si for silicate and X = C for carbonaceous grains)
		in both gas and dust phases,
		${\textstyle \mathcal{D}_{\rm{X}}}$ is the mass abundance of a dust species
		whose key element is X,
		$\textstyle\mathcal{R}$ is the returned fraction of the mass from stars formed, 
		$\textstyle\mathcal{Y}_{\rm{X}}$ is the mass fraction of element X that is newly produced and ejected by stars,
		${\textstyle f_{\rm{in, X}}}$ is the dust condensation efficiency of the element X in the ejecta, 
		$\psi$ is the star formation rate, 
		$\beta_\mathrm{SN}$ is the destruction efficiency of dust by SN shocks (defined below), and
		$[\mathrm{d}M_\mathrm{dust,X}/\mathrm{d}t]_\mathrm{acc}$ is the increasing rate of
		dust mass by accretion.
	The destruction efficiency by SN shock was introduced by \citet{mckee89} as
		$\textstyle\beta_{\rm{SN}}\equiv\epsilon_{\rm{s}}M_{\rm{s}}\gamma/\psi$,
		where $\textstyle\epsilon_{\rm{s}}$ is the fraction of dust destroyed in a
		single SN blast, and $\textstyle M_{\rm{s}}$ is the gas mass swept per
		SN blast, and $\gamma$ is the SN rate. We adopt $\beta_\mathrm{SN}=9.65$
		(HK11).
{In deriving equation (\ref{eq:dDdX}), some quantities are eliminated. The gas mass is used to normalize the metal and dust masses to obtain $Z_\mathrm{X}$ and $\mathcal{D}_\mathrm{X}$, respectively, and thus acts as a normalizing factor in our formulation. The star formation time-scale is also eliminated: changing the star formation time-scale just varies the quickness of metal and dust enrichment, leaving the relation between $Z_\mathrm{X}$ and $\mathcal{D}_\mathrm{X}$ unchanged. We refer to HK11 for the derivation of equation (\ref{eq:dDdX}). We assume $\mathcal{D}_\mathrm{X}=0$ at $Z_\mathrm{X}=0$.}

\subsection{Grain Growth by Accretion}

In HK11, we have formulated the grain growth by accretion considering the dependence on the grain size distribution. We also assume that the molecular clouds host both grain growth and star formation. Then,  we obtain (HK11)
	\begin{equation}
		\left[\frac{\mathrm{d}M_\mathrm{dust,X}}{\mathrm{d}t}\right]_{\rm{acc}}=\frac{\beta_\mathrm{X}\mathcal{D}_{\rm{X}}\psi}{\epsilon},\label{eq:dmdt}
	\end{equation}
	where $\textstyle\beta_{\mathrm{X}}$ is the mass increment of dust whose key
	species is X and $\epsilon$ is the star formation efficiency in a single molecular
	cloud. By assuming that grain growth is governed by the sticking of the key element X
	(i.e.\ Si for silicate and C for carbonaceous dust), HK11 derived the following formula
	for $\beta_\mathrm{X}$:
	\begin{equation}
		\beta_\mathrm{X}\simeq\left[\frac{\langle a^3\rangle_0}{3y\langle a^2\rangle_0+3y^2\langle a\rangle_0+y^3}+\frac{1-\xi_\mathrm{X}}{\xi_\mathrm{X}}\right]^{-1},\label{eq:beta}
	\end{equation}
where $y\equiv a_{0}\xi_\mathrm{X}\tau_{\rm{cl}}/\tau$
({$a_0=0.1~\micron$ is a typical grain radius,}
$\tau_\mathrm{cl}$ is the lifetime of molecular clouds, and
$\tau\propto a_{0}$ is the grain growth time-scale given
below in equation \ref{eq:tau}; {note that $a_0$ cancels
out in obtaining $y$ so that we can take an arbitrary value for
$a_0$}),
	$\langle a^\ell\rangle_0$
	is  the average of $a^\ell$ ($\ell =1$, 2, and 3) weighted for the grain
	size distribution (i.e.\ moments), and
	$\xi_\mathrm{X}=1-\frac{f_\mathrm{X}\mathcal{D}_\mathrm{X}}{Z_\mathrm{X}}$ is
	the fraction of element X in gas phase ($f_{\rm{X}}$ is the mass fraction of element
	X contained in the grains). The grain growth time-scale $\tau$ is given by
	\begin{equation}
		\tau\equiv a_0 \left / \left[\frac{m_{\rm{X}}S_{\rm{X}}}{f_{\rm{X}}\rho_{\rm{X}}}\left(\frac{Z_{\rm{X}}}{Z_{\rm{X}\sun}}\right)\left(\frac{\rm{X}}{\rm{H}}\right)_{\sun}n_{\rm{H}}\left(\frac{k_{\rm{B}}T_{\rm{gas}}}{2\pi m_{\rm{X}}}\right)^{\frac{1}{2}}\right] \right. ,\label{eq:tau}
	\end{equation}
	where 
		{$n_{\rm{H}}$ is the number density of hydrogen nuclei in the molecular clouds,} 
		$(\rm{X}/\rm{H})_{\sun}$ is the solar abundance of element $\rm{X}$ relative to hydrogen in number, 
		$\rho_{\rm{X}}$ is the material density of a dust grain whose key species is X, 
		$S_{\rm{X}}$ is the sticking probability of element $\rm{X}$ onto the dust,
		$m_{\rm{X}}$ is the atomic mass of element X,
		{$T_{\rm{gas}}$ is the gas temperature in the molecular clouds,} and
		$k_{\rm{B}}$ is the Boltzmann constant. 
	By using equation (\ref{eq:dmdt}), equation (\ref{eq:dDdX}) can be rewritten as
	\begin{equation}
		\mathcal{Y}_{\rm{X}}\frac{\mathrm{d}\mathcal{D}_{\rm{X}}}{\mathrm{d}Z_{\rm{X}}}=
		f_{\rm{in, X}}(\mathcal{R}Z_{\rm{X}}+\mathcal{Y}_{\rm{X}})
		-(\beta_{\rm{SN}}+\mathcal{R}-\beta_{\mathrm{X}}/\epsilon)\mathcal{D}_{\rm{X}}.
		\label{eq:dg_metal}
	\end{equation}

\subsection{Grain size distribution}

We assume that the dust and metals are predominantly supplied from SNe in high-$z$ ($z>5$) quasars {because they are the first significant dust source (Introduction). Although AGB stars may contribute to the dust enrichment even at $z>5$, our conclusion in this paper is not altered as long as AGB stars supply large ($\sim 0.1~\micron$) grains as suggested by observations \citep{groenewegen97,gauger99}.} The dust and metal yields are represented by the case of 20 M$_{\sun}$ progenitor with an unmixed helium core, according to \citet{hirashita10} (dust and metal yields are taken from \citealt{nozawa07} and \citealt{umeda02}, respectively). The grain species formed by SNe are C, Si, $\rm{SiO_2}$, Fe, FeS, $\rm{Al_2SO_3}$, MgO, $\rm{MgSiO_3}$ and $\rm{Mg_2SiO_4}$. Since the chemical properties of these species in grain growth is poorly known, we simply categorize them into two groups following \citet{hirashita10}: the carbonaceous species and the silicate species (all the species other than C; {Si is dominated}), and hypothesize that they evolve separately and independently in molecular clouds.

	\citet{nozawa07} have calculated the grain size distribution by considering dust condensation and destruction in a SN, {and have shown that small grains are efficiently trapped in the shocked region and considerably destroyed by thermal sputtering,} and thus the size distribution of grains finally ejected into the ISM is biased towards larger sizes. 
Because destruction of small grains with $a\la\mbox{a few}\times 10^{-2}~\micron$
is common {regardless of the progenitor mass and the treatment of mixing in the He core} in \citet{nozawa07} {(similar grain size distributions are also acquired by \citealt{bianchi07})}, it is sufficient to examine a single representative case for
the grain size distribution for SN-dust. 
This case is called `grain size distribution \textit{without} shattering'.

	\citet{hirashita10} have suggested that grains are processed by shattering after being injected into the ISM. Production of small grains by shattering increases the total surface of grains, which is expected to raise the grain growth efficiency. We adopt the grain size distribution after shattering from \citet{hirashita10} (the solar-metallicity case with hydrogen number density $n_\mathrm{H}=1$ cm$^{-3}$ and shattering duration $t=5$ Myr).
This case is called `grain size distribution \textit{with} shattering'.
Although it is difficult to constrain $t$ and $n_\mathrm{H}$
from observations, this single case is sufficient to demonstrate
the importance of the change of grain size distribution by shattering. {Since we only focus on grain growth around $Z\ga Z_{\sun}$, the assumption of solar metallicity is reasonable in this Letter.}

\begin{table*}
 \centering
 \begin{minipage}{140mm}
  \caption{Moments.}
  \label{tab:moment}
  \begin{tabular}{@{}lccccccrlr@{}}
  \toprule[1.6pt]
   & & & Silicate & & & & Carbonaceous & \\
 \cmidrule[0.8pt](lr){3-5}
 \cmidrule[0.8pt](lr){7-9} 
   & 
& $\langle a\rangle_{0}$ $(\rm{\mu m})$
& $\langle a^2\rangle_{0}$ $(\rm{\mu m}^2)$
& $\langle a^3\rangle_{0}$ $(\rm{\mu m}^3)$
& & $\langle a\rangle_{0}$ $(\rm{\mu m})$
& $\langle a^2\rangle_{0}$ $(\rm{\mu m}^2)$
& $\langle a^3\rangle_{0}$ $(\rm{\mu m}^3)$\\
  \midrule[0.8pt]
   Nozawa et al. (2007) & & $2.32\times 10^{-2}$ & $1.95\times 10^{-3}$ & $4.83\times 10^{-4}$ & & $8.59\times 10^{-3}$ & $3.45\times 10^{-4}$ & $6.35\times 10^{-5}$\\
   Hirashita et al. (2010) & & $7.84\times 10^{-4}$ & $3.23\times 10^{-6}$ & $4.35\times 10^{-7}$ & & $5.42\times 10^{-4}$ & $5.56\times 10^{-7}$ & $1.60\times 10^{-8}$\\
  \bottomrule[1.6pt]
\end{tabular}
\end{minipage}
\end{table*}

\begin{figure}
\includegraphics[width=0.45\textwidth]{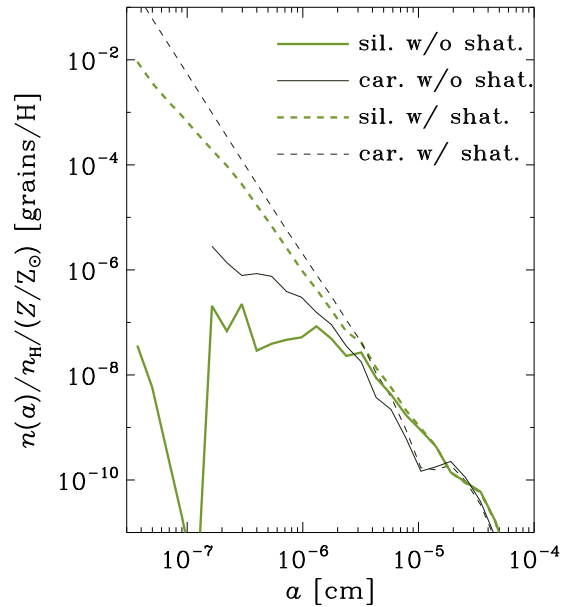}
 \caption{Grain size distributions adopted in this paper.
 {The grain size distributions are scaled with the number
 density of hydrogen nuclei and the metallicity.}
The solid and dashed lines show the size distributions
without and with shattering, respectively. The thick and
thin lines represent silicate and carbonaceous species,
respectively.}
 \label{fig:size}
\end{figure}

The above two grain size distributions adopted are shown in
Fig.\ \ref{fig:size} for silicate and carbonaceous species.
The moments of grain radii ($\langle a^\ell\rangle_0$)
necessary to estimate $\beta_\mathrm{X}$ in
equation (\ref{eq:beta}) are
listed in Table \ref{tab:moment}. We observe that
shattering makes the typical grain size smaller.
We fix $\langle a^\ell\rangle_0$; in other words,
we fix the grain size distribution throughout the
galaxy evolution to focus on the effects of
grain size distribution on grain growth.

\subsection{Selection of parameter values}

\begin{figure}
\includegraphics[width=0.4\textwidth]{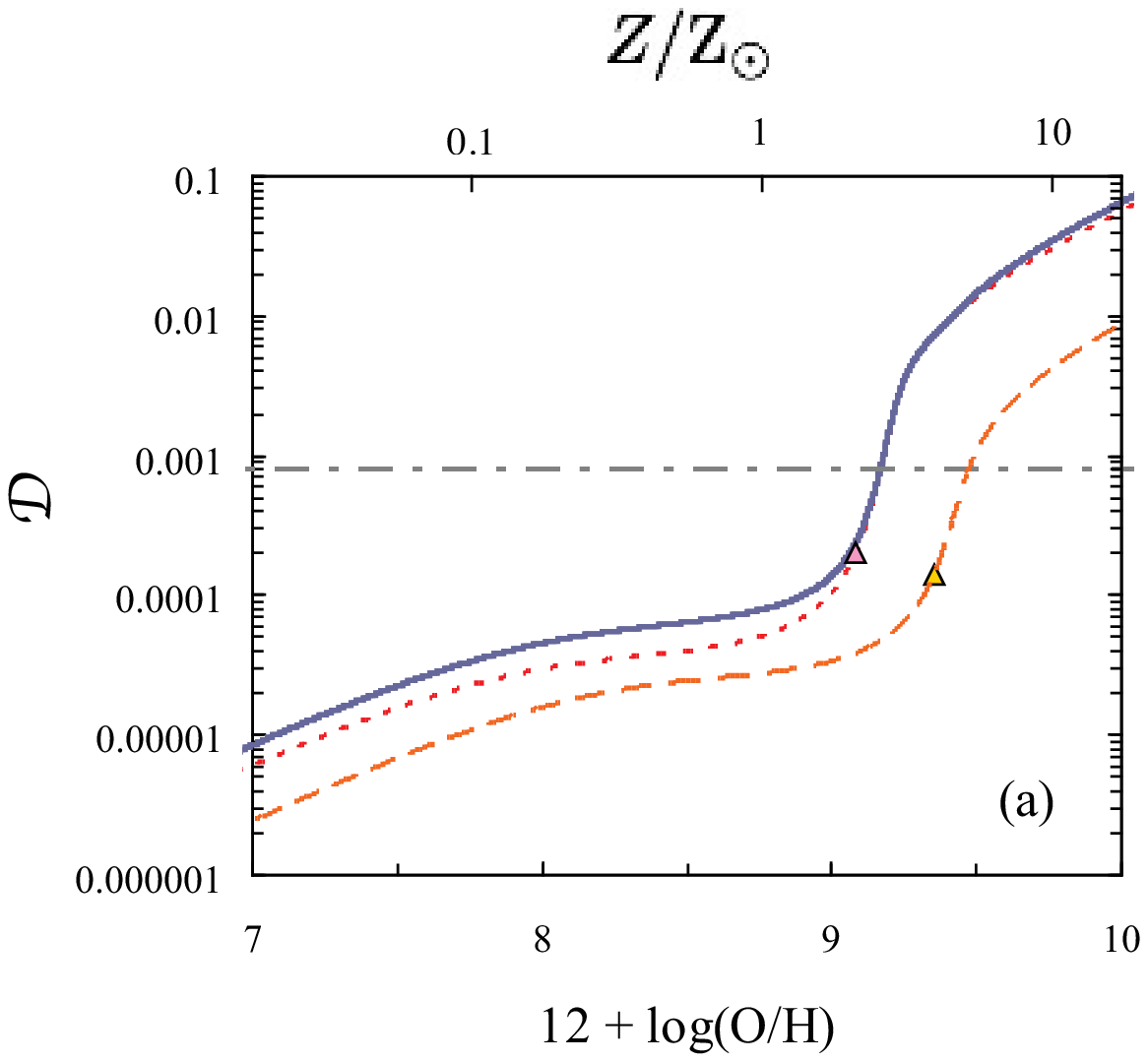}
\includegraphics[width=0.4\textwidth]{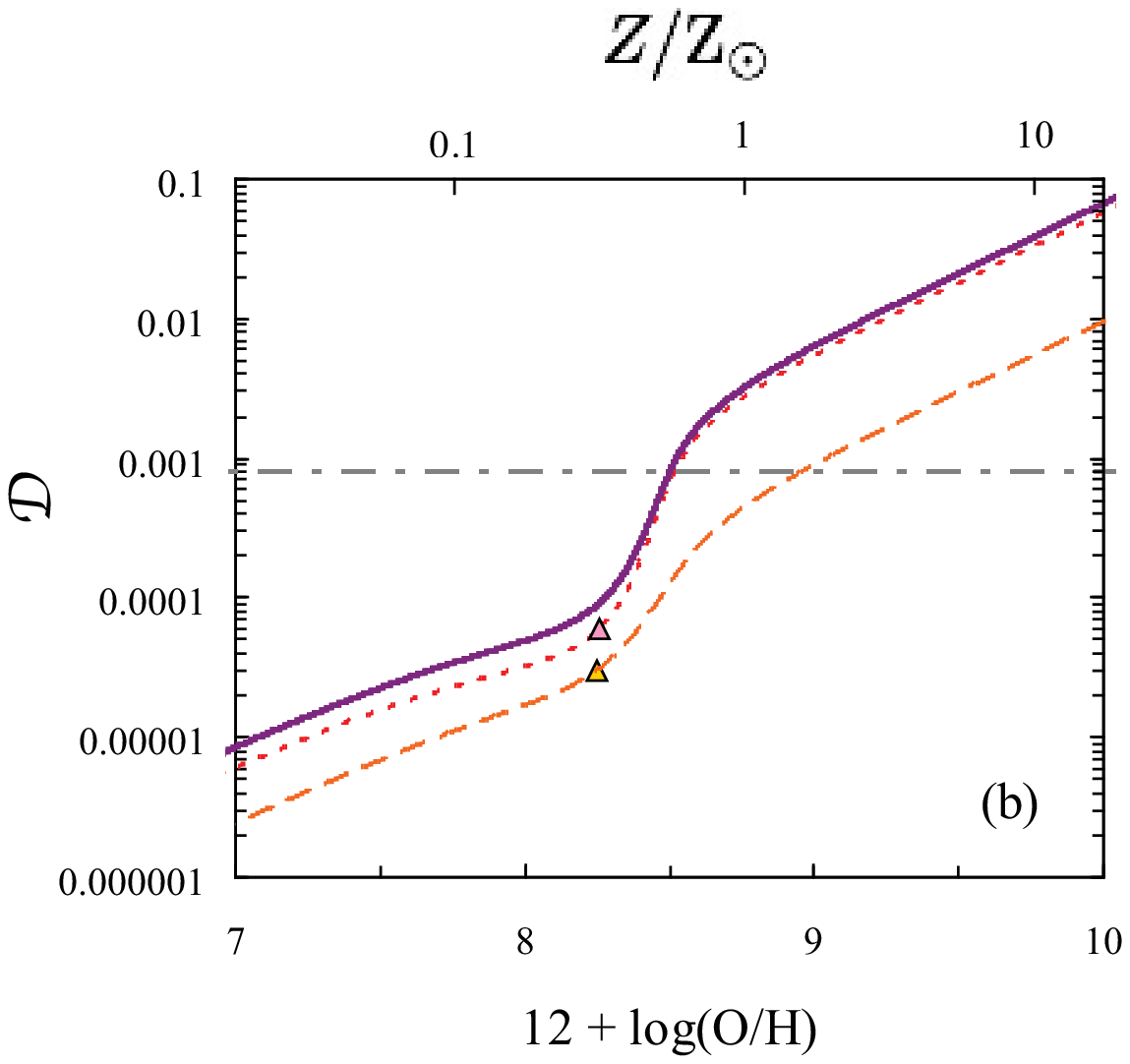}
\includegraphics[width=0.4\textwidth]{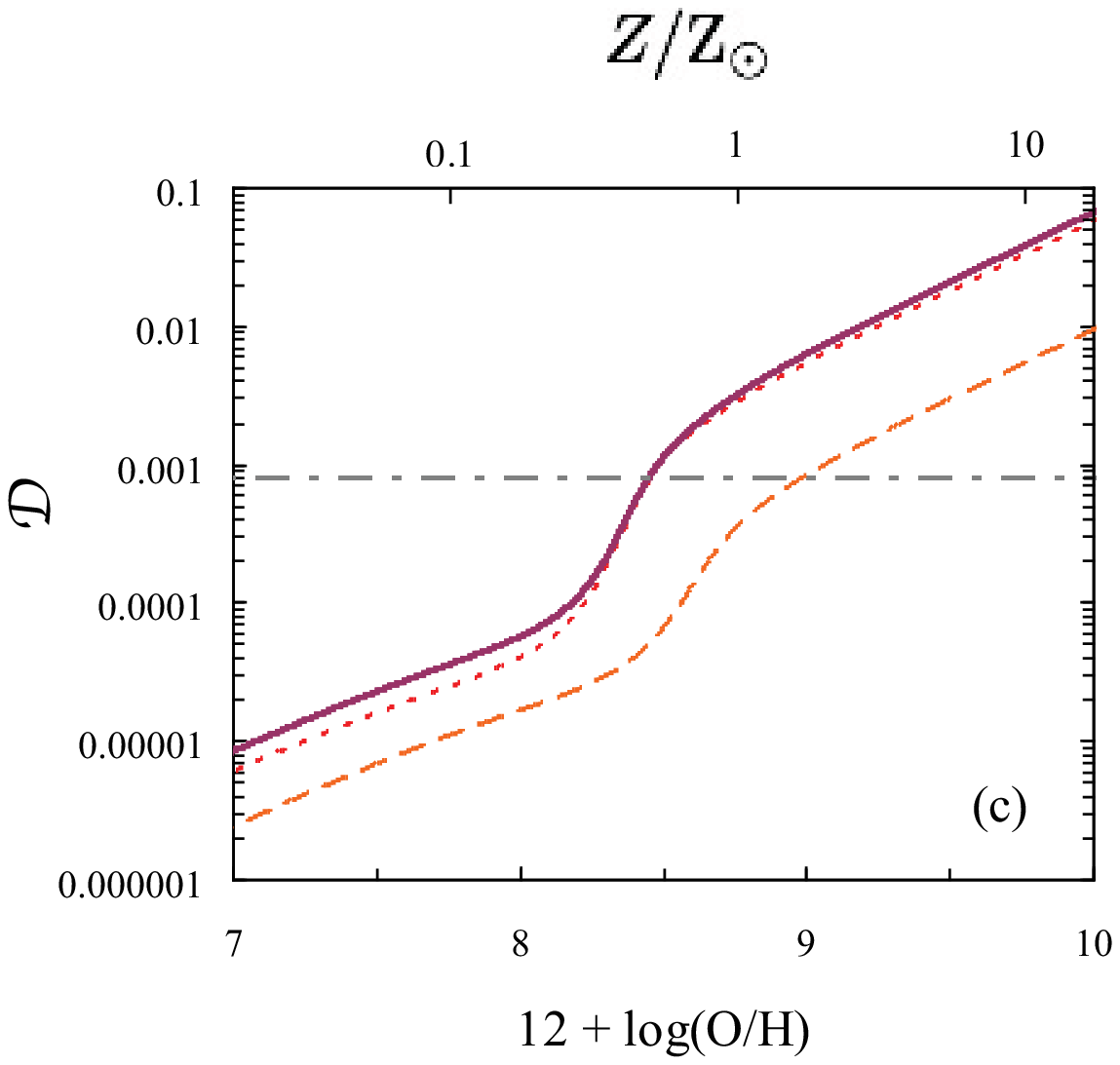}
 \caption{Relation between dust-to-gas ratio and metallicity.
 The solid lines shows the total dust-to-gas ratio. The dotted
 and dashed lines present the contributions from silicate and
 carbonaceous dust, respectively. The horizontal
dot-dashed line show the lower limit for the dust-to-gas
ratio in the high-$z$ quasar sample in \citet{michalowski10}.
Panels (a) and (b) present the results for the grain size distribution without and with shattering, respectively, {along with triangles marking the critical metallicities for grain growth.} Panel (c) shows the result for a
high density ($n_\mathrm{H}=10^4$ cm$^{-3}$).
 }
 \label{fig:dg_metal}
\end{figure}

 The typical time-scale of grain growth in
equation (\ref{eq:tau}) is estimated as (HK11): 
\begin{equation}
\tau=6.30\times 10^7 a_{0.1}\left(
Z_{\rm{Si}}/Z_{\mathrm{Si},\sun}\right)^{-1}
{n_3}^{-1}{T_{50}}^{-1/2}{S_{0.3}}^{-1}~\mathrm{yr}
\label{eq:tau_sil}
\end{equation}
for silicate, and
\begin{equation}
\tau=5.59\times 10^7 a_{0.1}\left(
Z_{\rm{C}}/Z_{\mathrm{C},\sun}\right)^{-1}
{n_3}^{-1}{T_{50}}^{-1/2}{S_{0.3}}^{-1}~\mathrm{yr}
\label{eq:tau_car}
\end{equation}
for carbonaceous species, where
	$a_{0.1}\equiv a_0/0.1\,\micron$, 
	$n_3\equiv n_{\rm{H}}/10^{3}\,\rm{cm^{-3}}$, 
	$T_{50}\equiv T_{\rm{gas}}/50\,\rm{K}$, and 
	$S_{0.3}\equiv S_{\mathrm{X}}/0.3$.
We {adopt the same values as HK11 (see also the
references therein):} 
	$n_{\rm{H}}=10^3\,\rm{cm}^{-3}$, $T_{\rm{gas}}=50\,\rm{K}$,
and $S_{\mathrm{X}}=0.3$. For the solar abundance of Si and C, we adopt (X/H)$_{\sun}$
listed in Table \ref{tab:parameter}, and convert it to $Z_\mathrm{X}$
by $Z_\mathrm{X}=(m_\mathrm{X}/\mu m_\mathrm{H})(\mathrm{X/H})$
($\mu =1.4$ is the correction for helium). We
adopt X, $f_\mathrm{X}$, $m_\mathrm{X}$, (X/H)$_{\sun}$, and $\rho_{\mathrm{X}}$
following HK11 (listed in Table \ref{tab:parameter}).
As a metallicity indicator, we adopt the oxygen abundance (O/H),
whose solar value is assumed to be $12+\log (\mathrm{O/H})=8.76$
\citep{lodders03}. The abundance of X is converted to that of O by
assuming the abundance pattern
calculated by the 20 M$_{\sun}$ SN model by \citet{umeda02}
(the ejected masses of O, C, and Si are 1.56, 0.257, and 0.257 M$_{\sun}$,
respectively).

In the SN model adopted \citep{umeda02,nozawa07},
$\textstyle{\mathcal{R}=8.87\times 10^{-2}}$, $\mathcal{Y}_{\rm{Si}}=9.22\times 10^{-4}$,
$\mathcal{Y}_{\rm{C}}=1.71\times 10^{-3}$, $\textstyle f_{\rm{in},Si}=0.351$
and $\textstyle f_{\rm{in},C}=0.135$. Besides, for the lifetime of a molecular
cloud, we adopt $\tau_\mathrm{cl}=10^7$ yr  \citep[e.g.][]{fukui10},
but we also mention a possibility of a longer lifetime in
Section \ref{subsec:mc}.

	\begin{table}
	\centering
		\begin{minipage}{140mm}
		\caption{Adopted quantities.}
	\label{tab:parameter}
			\begin{tabular}{@{}lccccccrlr@{}}
			\toprule[1.6pt]
				Species & X & $f_{\rm{X}}$ & $m_{\rm{X}}$ (amu)& $\textstyle\left(\rm{X}/\rm{H}\right)_{\sun}$ & $\rho_{\mathrm{X}}$ $\textstyle(\rm{g}/\rm{cm}^{3})$\\
			\midrule[0.8pt]
				Silicate & Si & 0.166 & 28.1 & $4.07\times 10^{-5}$ & 3.3 \\
				Carbonaceous & C & 1 & 12 & $2.88\times 10^{-4}$ & 2.26 \\
			\bottomrule[1.6pt]
			\end{tabular}
		\end{minipage}
	\end{table}

\section{Results}

\subsection{Without Shattering}

We examine the relation between dust-to-gas ratio and metallicity for
the grain size distribution without shattering. The result is exhibited in
Fig.\ \ref{fig:dg_metal}a. At low metallicities [$12+\log (\rm{O}/\rm{H})<8.5$],
the dust is simply supplied from SNe, which leads to an approximately
linear increase of the dust-to-gas ratio in terms of the metallicity.
{When $\mathcal{D}_\mathrm{X}$ reaches
$\sim f_\mathrm{in,X}\mathcal{Y}_\mathrm{X}/\beta_\mathrm{SN}$
($\sim 3\times 10^5$ and $2\times 10^{-5}$ for silicate and
carbonaceous dust, respectively), the dust destruction becomes
comparable to the dust supply from SNe (equation \ref{eq:dg_metal}).
Thus, the increase of the dust-to-gas ratio saturates around
$12+\log (\rm{O}/\rm{H})\simeq 8.5$.}
Around $12+\log (\rm{O}/\rm{H})\sim 9$ ($\sim 2$ Z$_{\sun}$) the dust-to-gas
ratio rapidly increases {because of the nonlinearity of grain growth
($\mathrm{d}\mathcal{D}_\mathrm{X}/\mathrm{d}Z_\mathrm{X}
\propto \mathcal{D}_\mathrm{X}Z_\mathrm{X}$; that is, grain growth
occurs as a result of collisions between dust grains and metals) in molecular clouds.
As shown in HK11, the grain growth becomes prominent at a certain
critical metallicity depending on dust grain size distribution
(see Section \ref{subsec:critical}).} Under the size distribution without shattering, dust growth becomes prominent only after the metallicity reaches a super-solar value. This is because of inefficient grain growth caused by the large surface-to-volume ratio of dust grains.

	In addition, SNe produce a large amount of silicate than carbonaceous dust. After grain growth in molecular clouds, the silicate species is still dominant over the carbonaceous species.


\subsection{With Shattering}

If we adopt the grain size distribution with shattering, we obtain the result exhibited in Fig.\ \ref{fig:dg_metal}b. Compared with the case in the previous subsection, the rapid increase of dust-to-gas ratio occurs at a lower metallicity, $12+\log (\rm{O}/\rm{H})\simeq 8.3$ ($\simeq 1/3$ Z$_{\sun}$), because shattering remarkably raises the surface-to-volume ratio of the dust grains.


\subsection{Observational Data}

\citet{michalowski10} have estimated the {dust and gas mass} by using the millimeter continuum and CO lines, respectively, for nine $z>5$ quasars. {Their samples show high dust-to-gas ratios from $1.65\times 10^{-2}$ to $4.30\times 10^{-2}$, which could be regarded as upper limits because they did not take neutral hydrogen into account. If we alternatively utilize their dynamical mass as an upper limit for the gas mass, we then obtain $\textstyle\mathcal{D}>8.0\times 10^{-4}$ (assuming an inclination angle of 40$^\circ$) to be a lower limit of the dust-to-gas ratio.} \citet{juarez09} claim that high-$z$ ($z>5$) quasars are likely to have super-solar metallicities ($\la 7$ Z$_{\sun}$) based on observations of broad line regions. {Narrow line regions (NLRs) may trace the metal content of the entire QSO better, and \citet{matsuoka09} indicate that the NLRs in quasars at $1\la z \la 4$ have solar metallicities. Considering that the bright QSOs at $z>5$ are highly biased to the most luminous and metal-enriched systems, it is reasonable to assume that the metallicities of $z>5$ QSOs is $\ga Z_{\sun}$.} Comparing our results with the observational lower limit of dust-to-gas ratio (Fig.\ \ref{fig:dg_metal}), we find that at least $12+\log(\rm{O}/\rm{H})=9.2$ ($Z\sim 3$ Z$_{\sun}$) is required to explain the lower-limit of dust-to-gas ratio if we adopt the grain size distribution without shattering. Thus, a super-solar metallicity is strongly required to explain the rich dust content in high-redshift quasars. With shattering, however, the dust-to-gas ratio reaches the observational lower limit at a more moderate metallicity ($12+\log(\rm{O}/\rm{H})=8.6$; $Z\sim 0.7$ Z$_{\sun}$) because of the grain growth at a lower metallicity.

\section{Discussion}

\subsection{Critical metallicity for grain growth}
\label{subsec:critical}

\citet{inoue11} and \citet{asano12} have shown that the grain growth by accretion dominates the grain abundance if the metallicity is larger than a certain critical metallicity. According to HK11, the critical metallicity $Z_{\rm{cr}}$ is the metallicity such that $\beta_\mathrm{X}(Z_{\rm{cr}})=\epsilon\beta_{\rm{SN}}=0.965$. For the dust grain size distribution without shattering, the critical metallicity of the silicate species and carbonaceous species are $12+\log(\rm{O}/\rm{H})=9.08$ and $12+\log(\rm{O}/\rm{H})=9.35$ ($\sim 2$--4 Z$_{\sun}$), respectively. These metallicities actually correspond to the metallicity levels where a strong rise of dust-to-gas ratio occurs by grain growth (Fig.\ \ref{fig:dg_metal}).
In contrast, for the dust grain size distribution with shattering, the critical metallicity becomes $12+\log(\rm{O}/\rm{H})=8.26$ and $12+\log(\rm{O}/\rm{H})=8.25$ ($\sim 1/3$ Z$_{\sun}$ for silicate and carbonaceous species. The variation in the critical metallicity between the grain size distributions with and without shattering explains the difference of the metallicity level at which the dust-to-gas ratio is raised by grain growth (see Fig.\ \ref{fig:dg_metal}).

\subsection{Dependence on other parameters}
\label{subsec:mc}

According to equation (\ref{eq:tau}), the hydrogen number density
in molecular clouds, $n_{\rm{H}}$, is also an important quantity
that determines the time-scale of grain
growth, $\tau$. We have presumed that $n_\mathrm{H}=10^{3}~\mathrm{cm}^{-3}$.
However, in some environments such as circumnuclear starburst regions in high-$z$ quasars, the physical conditions of molecular clouds can be quite different. Some recent observations imply that the density in molecular clouds
in high-$z$ quasars is higher than that in nearby galaxies
{\citep{riechers07,gao07,klessen07}}. In order to
investigate a possibility of a larger $n_\mathrm{H}$, we adopt
$n_\mathrm{H}=10^4$ cm$^{-3}$. In Fig.\ \ref{fig:dg_metal}c, we show the
result for the dust grain size distribution without shattering. We observe that higher
gas density enhances grain growth and that the rapid increase of
dust-to-gas ratio occurs at a lower metallicity. Therefore, the grain size
distribution without shattering can be reconciled with the observed
dust-to-gas ratio in high-$z$ quasars if we assume such a high density as
$n_\mathrm{H}\ga 10^4$ cm$^{-3}$.

Equation (3) offers another possibility of enhancing grain growth. The dust mass increment $\textstyle\beta_{\rm{X}}$ is a function of $\textstyle\tau_{\rm{cl}}/\tau$. As a result, a longer lifetime of molecular clouds hosting the grain growth $\textstyle\tau_{\rm{cl}}$ has the same effect as a shorter typical time-scale of grain growth $\tau$, which means that $\textstyle\tau_{\rm{cl}}=10^8~\mathrm{yr}$ instead of $\textstyle\tau_{\rm{cl}}=10^7~\mathrm{yr}$ has the same effect as applying $n_\mathrm{H}=10^{4}~\mathrm{cm}^{-3}$ instead of $n_\mathrm{H}=10^{3}~\mathrm{cm}^{-3}$. {\citet{koda09} have indicated that the existence of sheared structure within the spiral arms in the Milky Way supports a lifetime of molecular clouds comparable to galactic rotational time-scales ($\sim 10^{8}~\rm{yr}$)}. Moreover, the gas temperature $T_{\rm{gas}}$ and the sticking probability $S_{\mathrm{X}}$ also enter the time-scale (see equations \ref{eq:tau_sil} and \ref{eq:tau_car}); however, since $T_{\mathrm{gas}}$ and $S_{\mathrm{X}}$ cannot be much larger than the adopted values, it is difficult to make $\tau$ shorter by changing $T_{\mathrm{gas}}$ or $S_{\mathrm{X}}$.

\section{Conclusion}

The importance of grain size distribution on dust enrichment in high-redshift ($z>5$) quasars has been investigated in this work. We have found that, if the grain size distribution of SN-dust is considered, very high metallicity ($\textstyle\ga 3Z_{\sun}$) is required to explain the dust-to-gas ratio of those quasars under a similar quantities to the local molecular clouds ($\textstyle n_{\rm{H}}=10^{3}$ cm$^{-3}$ and $\textstyle\tau_{\mathrm{cl}}=10^7$ yr). However, if small grains are produced by shattering, grain growth is activated and the dust-to-gas ratio in high-$z$ quasars can be explained by more moderate metallicities($\ga 0.7$ Z$_{\sun}$). Thus, we conclude that the dust grain size distribution has a dramatic impact on the evolution of dust content in high-$z$ quasars.

\section*{Acknowledgments}
{We thank T. Nozawa for providing us with the numerical
data for the yield of various dust species in supernova ejecta.}
H.H. is supported by NSC grant 99-2112-M-001-006-MY3.

\bsp

\label{lastpage}


\begin{thebibliography}{}
\bibitem[\protect\citeauthoryear{Asano et al.}{2012}]{asano12}
    Asano, R. S., Takeuchi, T. T., Hirashita, H., \& Inoue, A. K. 2012,
    Earth Planets Space, submitted
\bibitem[\protect\citeauthoryear{Bianchi \& Schneider}{2007}]{bianchi07}
    Bianchi, S., \& Schneider, R. 2007, MNRAS 378, 973
\bibitem[\protect\citeauthoryear{Draine}{2009}]{draine09}
    Draine, B. T. 2009, in Henning Th., Gr\"{u}n E., Steinacker J.,
    eds, Cosmic Dust -- Near and Far. ASP Conf.\ Ser., ASP,
    San Francisco, p.\ 453
\bibitem[\protect\citeauthoryear{Dwek}{1998}]{dwek98}
    Dwek, E. 1998, ApJ, 501, 643
\bibitem[\protect\citeauthoryear{Fukui \& Kawamura}{2010}]{fukui10}
    Fukui, Y., \& Kawamura, A. 2010, ARA\&A, 48, 547
\bibitem[\protect\citeauthoryear{Gall et al.}{2011}]{gall11}
    Gall, C., Andersen, A. C., \& Hjorth, J. 2011, A\&A, 528, A14
\bibitem[\protect\citeauthoryear{Gallerani et al.}{2010}]{gallerani10}
    Gallerani, S., et al.\ 2010, A\&A, 523, 85
\bibitem[\protect\citeauthoryear{Gao et al.}{2007}]{gao07}
    Gao, Y., Carilli, C. L., Solomon, P. M., \& Vanden Bout, P. A. 2007,
    ApJ, 660, L93
\bibitem[\protect\citeauthoryear{Gauger et al.}{1999}]{gauger99}
    Gauger, A., Balega, Y. Y., Irrgang, P., Osterbart, R., \& Weigelt, G. 1999,
    A\&A, 346, 505
\bibitem[\protect\citeauthoryear{Groenewegen}{1997}]{groenewegen97}
    Groenewegen, M. A. T. 1997, A\&A, 317, 503
\bibitem[\protect\citeauthoryear{Hirashita}{1999}]{hirashita99}
    Hirashita, H. 1999, ApJ, 510, L99
\bibitem[\protect\citeauthoryear{Hirashita \& Kuo}{2011}]{hirashita11}
    Hirashita, H., \& Kuo, T.-M. 2011, MNRAS 416, 1340 (HK11)
\bibitem[\protect\citeauthoryear{Hirashita et al.}{2010}]{hirashita10}
    Hirashita, H., Nozawa, T., Yan, H., \& Kozasa, T. 2010, MNRAS,
    404, 1437
\bibitem[\protect\citeauthoryear{Inoue}{2003}]{inoue03}
    Inoue, A. K. 2003, PASJ, 55, 901
\bibitem[\protect\citeauthoryear{Inoue}{2011}]{inoue11}
    Inoue, A. K. 2011, Earth Planets Space, 63, 1027
\bibitem[\protect\citeauthoryear{Juarez et al.}{2009}]{juarez09}
    Juarez, Y., Maiolino, R., Mujica, R., Pedani, M., Marinoni, S.,
    Nagao, T., Marconi, A., \& Oliva, E. 2009, A\&A, 494, L25
\bibitem[\protect\citeauthoryear{Klessen, Spaans, \& Jappsen}{2007}]{klessen07}
    Klessen, R. S., Spaans, M. \& Jappsen, A.-K. 2007, MNRAS 374, L29
\bibitem[\protect\citeauthoryear{Koda et al.}{2009}]{koda09}
Koda, J., et al.,
2009, ApJ, 400, L132
\bibitem[\protect\citeauthoryear{Lodders}{2003}]{lodders03}
    Lodders, K. 2003, ApJ, 591, 1220
\bibitem[\protect\citeauthoryear{Maiolino et al.}{2004}]{maiolino04}
    Maiolino, R., Schneider, R., Oliva, E., Bianchi, S., Ferrara, A.,
    Mannucci, F., Pedani, M., \& Roca Sogorb, M. 2004, Nature, 7008, 533
\bibitem[\protect\citeauthoryear{Mathis, Rumpl, \& Nordsieck}{1977}]{mathis77}
    Mathis, J. S., Rumpl, W., \& Nordsieck, K. H. 1977, ApJ, 217, 425
\bibitem[\protect\citeauthoryear{Matsuoka et al.}{2009}]{matsuoka09}
    Matsuoka, K., Nagao, T., Maiolino, R., Marconi, A., \&
    Taniguchi, Y. 2009, A\&A, 503, 721
\bibitem[\protect\citeauthoryear{Mattsson}{2011}]{mattsson11}
    Mattsson, L. 2011, MNRAS, 414, 781
\bibitem[\protect\citeauthoryear{McKee}{1989}]{mckee89}
    McKee, C. F. 1989, in Allamandola L. J. \& Tielens A. G. G. M. eds.,
    IAU Sump.\ 135, Interstellar Dust, Kluwer, Dordrecht, 431
\bibitem[\protect\citeauthoryear{Micha{\l}owski et al.}{2010}]{michalowski10}
    Micha{\l}owski, M. J., Murphy, E. J., Hjorth, J., Watson, D., Gall, C., \&
    Dunlop, J. S. 2010, A\&A, 522, A15
\bibitem[\protect\citeauthoryear{Nozawa et al.}{2007}]{nozawa07}
    Nozawa, T., Kozasa, T., Habe, A., Dwek, E., Umeda, H., Tominaga, N.,
    Maeda, K., \& Nomoto, K. 2007, ApJ, 666, 955
\bibitem[\protect\citeauthoryear{Pei}{1992}]{pei92}
    Pei, Y. C. 1992, ApJ, 395, 130
\bibitem[\protect\citeauthoryear{Pipino et al.}{2011}]{pipino11}
    Pipino, A., Fan, X. L., Matteucci, F, Calura, F., Silva, L.,
    Granato, G., \& Maiolino, R. 2011, A\&A, 525, A61
\bibitem[\protect\citeauthoryear{Riechers et al.}{2007}]{riechers07}
    Riechers, D. A., Walter, F., Carilli, C. L., \& Bertoldi, F. 2007, ApJ,
    671, L13
\bibitem[\protect\citeauthoryear{Umeda \& Nomoto}{2002}]{umeda02}
    Umeda, H., \& Nomoto, K. 2002, ApJ, 565, 385
\bibitem[\protect\citeauthoryear{Valiante et al.}{2009}]{valiante09}
    Valiante, R., Schneider, R., Bianchi, S., \& Andersen, A. C. 2009,
    MNRAS, 397, 1661
\bibitem[\protect\citeauthoryear{Valiante et al.}{2011}]{valiante11}
    Valiante, R., Schneider, R., Salvadori, S., \& Bianchi, S., 2011,
    MNRAS, 416, 1916
\bibitem[\protect\citeauthoryear{Yan, Lazarian, \& Draine}{Yan et al.}{2004}]{yan04}
    Yan, H., Lazarian, A., \& Draine, B. T. 2004, ApJ, 616, 895
\bibitem[\protect\citeauthoryear{Zhukovska, Gail, \& Trieloff}{Zhukovska et al.}{2008}]
{zhukovska08}
    Zhukovska, S., Gail, H.-P., \& Trieloff, M. 2008, A\&A, 479, 453
\end{thebibliography}
\end{document}